# Forced Yet Free:

## What Magicians' Forcing Reveals Beyond Intentional Binding

**TOIDA Koichi**
MESON, Inc.
koichi.toida@meson.tokyo

**ABSTRACT** In magicians' forcing, spectators may experience a choice as self-determined even when that choice has been externally directed. Research on the sense of agency has developed largely around action–outcome relations, most notably intentional binding; however, the problem of choice authorship—why a particular choice is experienced as originating from oneself—must be treated as distinct. This paper integrates research on agency and forcing by distinguishing the locus of intervention within the choice–action–outcome chain, self-attribution at the Decision, Action, and Outcome levels, and the processes by which feeling of agency and judgment of agency are constructed. On this basis, a distinction is drawn between intentional binding, which concerns the temporal/causal relation between action and outcome, and the different binding problem exposed by forcing: the relation between a choice and its author. This theoretical relation is termed authorship binding. Authorship binding does not denote a new implicit measure; rather, it refers to the constructive relation through which a choice whose formation has been directed by external factors can nevertheless be experienced as originating from the self. Forcing can sustain this relation not by eliminating agency, but by selectively preserving, substituting, and redistributing agency cues. XR, in which body-centred spatial relations can be manipulated, is further conceptualised as a medium for extending the forced-yet-free structure into space.

## I. INTRODUCTION: From Choice to Authorship

In mentalism, exemplified by number divination and card prediction, a performer predicts or identifies a number, card, word, or object immediately after a spectator has ostensibly chosen it for themselves. Here, mentalism refers to a genre of magic involving phenomena such as mind reading, clairvoyance, prediction, and thought insertion (Pailhès et al., 2023). The present analysis, however, does not concern mentalism as a whole, but magicians' forcing: techniques that offer spectators a choice while guiding them towards a particular choice or outcome. In forcing, a choice is presented as though it had been known or predicted in advance by the performer while the spectator retains the experience that they 'really chose it themselves' (Pailhès et al., 2020). Experiments using multiple forcing techniques have likewise shown that a subjective sense of free choice and agency can be preserved (Pailhès & Kuhn, 2020; Pailhès et al., 2021).

What makes this phenomenon theoretically interesting is that two facts can coexist: the choice has been causally influenced from outside, yet the spectator experiences it as 'my choice'. If the choice is experienced as wholly forced, the prediction becomes merely procedural; if the performer has merely guessed the outcome by chance, it does not appear to be mind reading. The effect of forcing therefore arises from preserving the spectator's experience of being the author of the choice while destabilising the available causal explanation of how that choice was formed.

This problem can be connected to research on the sense of agency (SoA). SoA generally refers to the subjective experience of initiating and controlling one's actions and of bringing about events in the external world through those actions (Gallagher, 2000; Haggard, 2017). Agency is not an introspective capacity for directly apprehending objective causal relations. Rather, it is constructed through the integration of multiple sources of information, including motor predictions, sensory feedback, action outcomes, prior beliefs, and contextual cues (Moore et al., 2009; Synofzik et al., 2008). It has also been shown that people are more likely to infer that they caused an outcome when a thought precedes the action and is consistent with the outcome, and when alternative causes are not salient (Wegner & Wheatley, 1999).

Intentional binding has played a particularly prominent role in research on agency. The perceived times of an intentional action and its outcome have been shown to shift towards one another (Haggard et al., 2002), and this phenomenon has been widely used as an implicit measure associated with SoA (Moore & Obhi, 2012). Subsequent research, however, has shown that temporal binding can also occur for mechanical causal relations (Buehner, 2012), and that binding does not always differ between voluntary action and passive movement (Kirsch et al., 2019; Kong et al., 2024). The temporal attraction phenomenon termed intentional binding has not itself been disconfirmed, but interpretations that regard it as specific to intentional action, or as a unique direct readout of SoA, have been reconsidered. A Bayesian causal-inference account has also been proposed, integrating a causal belief that action and outcome arise from a common cause with predictions about their temporal relation (Tanaka, 2024).

The critical point is that intentional binding primarily concerns the relation between action and outcome. It provides substantial information about the subjective construction of a causal sequence in which 'this action produced this outcome', but does not directly indicate why a particular action was chosen or whether the resulting choice is attributed to the self. Indeed, action choice can alter explicit agency ratings without altering temporal binding, suggesting that the two do not necessarily reflect the same information (Schwarz et al., 2019). Conversely, the number of available actions has been reported to affect binding (Barlas & Obhi, 2013), indicating that choice and binding are not wholly unrelated. These mixed findings suggest that, rather than identifying agency over choice with action–outcome binding, the relation between them must itself be examined.

Both prospective cues available before action and retrospective cues available after action contribute to the construction of agency. Fluency of action selection can contribute to judgments of agency independently of outcome monitoring (Sidarus et al., 2013, 2017). Moreover, even when nearly identical actions are performed, SoA can be retrospectively altered by outcome feedback indicating success or failure (Oishi et al., 2018). Experimental work manipulating goal achievement in combination with action control and outcome control has shown across multiple experiments that failure to attain a goal substantially reduces SoA (Saad et al., 2026). Agency is therefore not determined solely by the precision of action–outcome correspondence, but is constructed from multiple cues involving selection, outcome, goals, and context.

A related dissociation between lower-level sensory processing and explicit authorship is evident in studies of delayed auditory feedback. Introducing a delay into auditory feedback accompanying voluntary movement reduces the feeling of authorship over the sound (Toida et al., 2016). In that electroencephalographic study, two event-related potential (ERP) components, enhanced-P2 and N300, were identified in response to delayed auditory feedback. Enhanced-P2 was associated with the processing of temporal deviation for short delays around the threshold for conscious delay detection, whereas N300 was associated with delay magnitude, conscious detection of delay, and reduced authorship. Subsequent ERP studies have likewise shown that modulation of the auditory P2 is involved in discriminating self-generated from other-generated sounds and in action-timing control (Bolt & Loehr, 2021, 2023; Han et al., 2025), and P2 has been examined as a neural index of the processing of self-generated auditory consequences. These findings suggest that caution is required before treating low-level processing of self-generation, conscious mismatch detection, and explicit judgments of authorship as a single process.

The same decomposition is required for choice. The experience of having initiated and finalised a decision or intention oneself has been explicitly characterised as Decision-level Agency and distinguished from Action-level Agency over bodily movement and Outcome-level Agency over external consequences (Kong, 2026). Once this distinction is adopted, two different binding problems become apparent within agency research. One concerns how action and outcome are bound into the same causal sequence. The other concerns how a choice is bound to the self—that is, to the author of the decision.

The latter is termed authorship binding. This term does not denote a temporal-perceptual phenomenon intended to replace intentional binding, nor does it introduce a new implicit measure. Authorship binding refers to the constructive relation through which a choice is attributed to the self as one's own decision, even when its formation has been directed by external causal factors. Whereas intentional binding concerns the subjective temporal/causal coupling of action and outcome, authorship binding concerns the subjective attribution of a decision to its author.

It is here that magicians' forcing becomes theoretically informative. Forcing comprises techniques that lead spectators towards a particular choice or outcome while preserving the experience that they have been given a choice (Pailhès et al., 2020). It is not merely a selection bias. Across the process by which spectators compare, choose, act, and receive an outcome, forcing manipulates which agency cues are preserved, which causal relations are severed, and which interpretations can be established retrospectively. It therefore provides a boundary case in which authorship binding may be preserved despite objective intervention in choice formation.

The question addressed in this paper is: how does forcing construct agency? The argument does not reduce magic performance as a whole to agency. Rather, forcing is treated as a structure that intervenes at different points in the choice–action–outcome chain while selectively preserving other agency cues. XR is further conceptualised not merely as an apparatus for forcing experiments, but as a medium through which choice formation and choice authorship can be dissociated by manipulating body-centred spatial relations.

The paper makes three theoretical contributions. First, it distinguishes the action–outcome relation primarily examined through intentional binding from the authorship relation by which a choice is attributed to the self, thereby describing forced yet free as a problem of choice authorship. Second, it locates forcing techniques that operate on selection, outcome, and interpretation as distinct interventions along the choice–action–outcome chain and analyses which forms of Decision-, Action-, and Outcome-level Agency are preserved or disrupted. Third, it extends this framework to XR and argues that body-centred spatial relations—including distance, direction, reachability, and motor cost—can bias choice probabilities without eliminating the available options.

## II. Agency Across Choice–Action–Outcome

### II.I. Targets, Levels, and Construction of Agency

Research on agency has shown that self-experience is not a direct reflection of objective bodily states or causal relations. Agency and body ownership are dissociable: the experience that 'this body is mine' is not identical to the experience that 'I am causing this movement' (Gallagher, 2000; Tsakiris et al., 2007). Efferent information also plays a specific role in recognising one's own actions (Tsakiris et al., 2005). Self-attribution is therefore not based on a single signal, but is constructed from multiple sources of information.

A distinction can be drawn between a pre-reflective, low-level feeling of agency—the feeling of being the agent—and a judgment of agency, in which sensorimotor cues are integrated with context, beliefs, and inference to explicitly attribute agency to oneself or another person (Synofzik et al., 2008; Haggard & Tsakiris, 2009). These two processes are related but not identical. Smooth sensorimotor correspondence tends to preserve a local feeling of agency, whereas judgments about why a particular action or choice occurred can vary as a function of prospective cues, outcomes, contextual information, and retrospective inference.

The same decomposition is required within the choice–action–outcome chain. Decision-level, Action-level, and Outcome-level Agency ordinarily covary but are not identical (Kong, 2026). If, after being instructed by an experimenter to choose object A, a participant grasps A with their own hand and produces the intended consequence, Action-level and Outcome-level Agency may remain high even though Decision-level Agency is low. Conversely, if the participant chooses A themselves but only the trajectory of a virtual hand is automatically corrected, Decision-level Agency may be preserved while Action-level Agency is weakened. If the participant makes both the choice and the action themselves, but the outcome is determined independently of both, only Outcome-level Agency may be impaired.

The three levels adopted here are not identical to Pacherie's (2008) hierarchy of intentions comprising distal, proximal, and motor intentions. Whereas that hierarchy organises functional levels of planning and execution leading to action, the present distinction describes whether a decision, bodily action, or external outcome is attributed to the self. This distinction implies that agency at one level cannot be directly inferred from a measure of agency at another. Intentional binding concerns the temporal/causal relation between action and outcome, but does not directly indicate why a particular action was chosen. Motor–sensory attenuation likewise primarily concerns the relation between action and sensory consequence. A target-choice rate can show that the environment influenced the distribution of choices, but cannot establish whether the resulting choice was experienced as originating from the self. Explicit authorship, temporal binding, and behavioural choice are not simple proxy measures for one another.

This point is also evident from studies that manipulate action choice. Although one study reported that a larger number of available options increased intentional binding (Barlas & Obhi, 2013), another found that action choice altered explicit agency ratings without altering temporal binding (Schwarz et al., 2019). Thus, self-attribution concerning choice and action–outcome temporal binding may be related, but they are not the same phenomenon.

### II.II. Predictive and Retrospective Agency

In the comparator model, agency is supported by a match between the predicted consequences of an action and the actual sensory feedback (Frith et al., 2000). This predictive mechanism is also reflected in the attenuation of self-generated stimuli. When movement of one hand, transmitted through two robotic arms, is used to stimulate the opposite palm, the stimulation is perceived as more intense as the temporal or spatial mismatch between the movement and tactile stimulation increases, indicating that self-generated sensory attenuation depends on the correspondence between prediction and feedback (Blakemore et al., 1999). Moreover, when a sustained delay is introduced into auditory feedback accompanying self-generated movement, subjective simultaneity is recalibrated towards the experienced delay distribution (Toida et al., 2012, 2014). This shows that the evaluation of action–outcome timing is not fixed, but is updated according to the temporal statistics of the environment.

Likewise, a simple match between prediction and feedback is insufficient to account for intentional binding. A Bayesian causal-inference model integrates a prior belief that action and outcome arise from a common cause, the observed time difference, the uncertainty of each signal, and a temporal prediction of how quickly a causally related outcome should occur. As the inferred probability of a common cause increases, the perceived times of action and outcome are reconstructed so as to fit the same causal sequence, compressing the subjective interval. Computational comparisons have shown that this framework can account for temporal estimates in multiple observers (Tanaka, 2024). However, not all observers follow a single Bayesian strategy, and this should not be taken as the sole established mechanism of intentional binding.

Agency, however, is not determined by prediction error alone. The fluency of action selection constitutes a prospective cue that can contribute to SoA independently of actual outcome monitoring (Sidarus et al., 2013, 2017). Conversely, even when the action process is nearly identical, judgments of agency can be altered retrospectively by outcome feedback indicating success or failure (Oishi et al., 2018). Goal achievement has likewise been shown to contribute strongly to SoA when goals, action control, and outcome control are systematically manipulated (Saad et al., 2026).

Accordingly, the experience that 'I caused this' must be understood as the integration of multiple cues, including not only action–feedback correspondence, but also the fluency of selection preceding the action, consistency with the outcome, goal achievement, and the plausibility of a causal explanation. Multiple components must likewise be distinguished in the subjective experience of choice. Decision authorship is the experience of having initiated and settled the decision oneself. Perceived freedom is the experience that one could have selected another option. Perceived external influence is the experience of having been directed by the environment or another person. 'I decided', 'I could have chosen something else', and 'I was not guided' may correlate, but they are not synonymous.

## II.III. From Intentional Binding to Authorship Binding

Intentional binding was initially described as a phenomenon that binds action and outcome within subjective time. Subsequent research has shown that this binding depends not only on intentionality, but also on causal relations, temporal prediction, beliefs, and signal precision (Buehner, 2012; Kirsch et al., 2019; Kong et al., 2024; Tanaka, 2024). The theoretical value of intentional binding therefore lies not so much in a 'direct readout of intention' as in its capacity to probe how action and outcome are constructed as belonging to a single causal sequence. Previous work has also explicitly related intentional binding to personal authorship over action outcomes. Ebert and Wegner (2010) showed that action–event consistency increased both temporal binding and self-reported authorship, while also finding evidence that the two measures were dissociable. Importantly, however, the authorship examined in that work concerned whether one's action caused a subsequent event. The present account addresses a different relation: whether the choice itself is attributed to the self as its author.

This distinction becomes critical when choice formation itself is externally directed. Even when a choice is enacted through one's own bodily movement, the formation of that choice may have been influenced by external factors. The relevant question is then not 'did this action produce the outcome?' but 'did this decision originate from me?'

The relation between a choice and the self as its subjective author is termed authorship binding. Authorship binding does not imply temporal compression in perception, nor does it presuppose the same measurement paradigm as intentional binding. The term binding is used to capture a theoretical commonality: independently of the objective causal sequence, multiple events or representations can be linked into a single structure of self-attribution.

Decision-level Agency and authorship binding are related but not synonymous. Decision-level Agency identifies the level at which agency is experienced—that is, whether the initiation and settlement of a decision are attributed to the self. Authorship binding, by contrast, identifies the self-attributive relation to be explained at that level: the relation through which a particular choice is experienced as one's own. The former therefore specifies a level of agency, whereas the latter specifies the relation between a choice and its subjective author. This choice-specific attribution is distinct from the more fundamental for-me-ness of experience—its first-personal givenness irrespective of whether a particular choice is judged to be one's own—as examined in *Present After Presence* (Toida, 2026a).

Whereas intentional binding primarily concerns the Action–Outcome link, authorship binding concerns the Self–Choice link. The central question for the former is, 'Did my action produce this effect?' The question for the latter is, 'Did this choice originate from me?' In ordinary voluntary action these two relations tend to coincide, because one forms the choice oneself, executes the action oneself, and the action produces the outcome. Forcing, however, partially severs this chain. Self–Choice attribution can be preserved even when choice probability is externally manipulated, and judgments of control can be preserved even when Action–Outcome contingency is severed. Forcing is therefore a boundary case that dissociates multiple binding relations that ordinarily covary.

## II.IV. The Forced-yet-Free Structure

Human choices are not made in a vacuum. They are continually influenced by decision information and decision structure, including the spatial arrangement of options, presentation order, salience, defaults, and social information (Johnson et al., 2012; Münscher et al., 2016). If agency were defined as absent whenever an external factor changed choice probability, almost every choice would become non-agentic. Conversely, a subjective report that 'I decided for myself' is not evidence that external influence was absent.

From an affordance perspective, an affordance is an opportunity for action that the environment provides in relation to an actor's bodily capacities, and it cannot be reduced either to a property of the environment alone or to the actor's subjective state alone (Gibson, 1979). It has also been proposed that multiple action candidates are specified in parallel and that competition among them is biased by bodily state, goals, spatial arrangement, and related factors (Cisek, 2007). From this perspective, a selection force can be described as a state in which the action affordance associated with one option is stronger than those associated with the alternatives, thereby biasing choice probability without eliminating the alternatives.

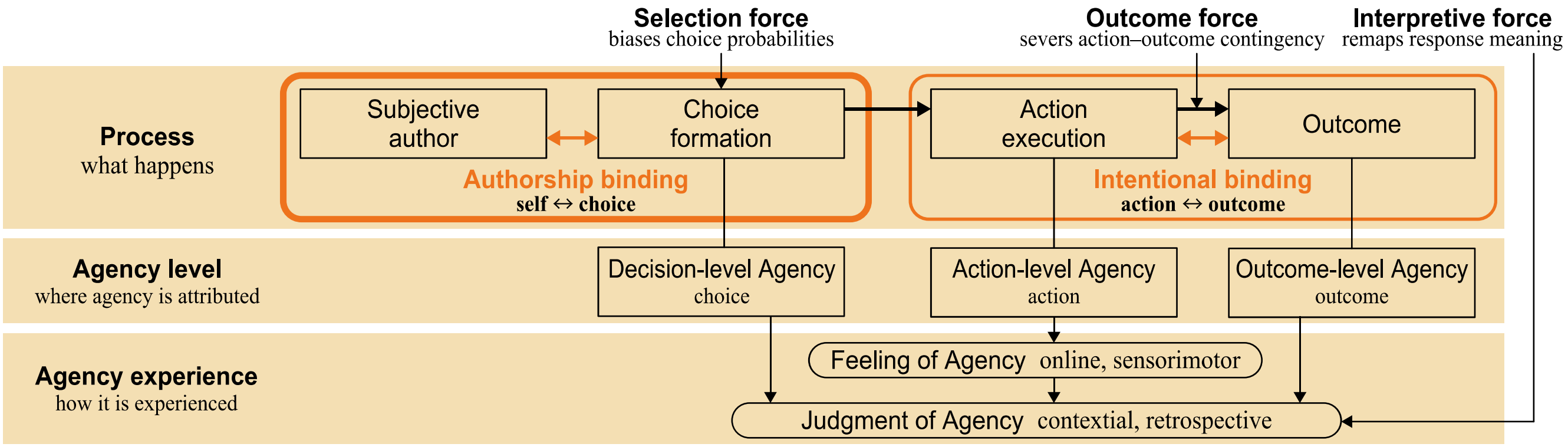

**FIGURE 1** Conceptual framework of forcing and agency. The model distinguishes objective interventions along the choice–action–outcome process, the level at which agency is attributed, and the mode in which agency is experienced. Authorship binding links the subjective author to the resulting choice, whereas intentional binding links action execution to outcome. Selection, outcome, and interpretive forces intervene at different points while potentially preserving subjective authorship, giving rise to the forced-yet-free structure.

Forced yet free refers to a state in which authorship binding is preserved despite external causal influence. This is not a logical contradiction. 'The choice was influenced' is a third-person description of the causes that altered the distribution of choices, whereas 'I chose' is a first-person description of the subjective authorship of a choice. The theoretical value of forcing is not that it adjudicates whether free will exists. Rather, it permits a separation between the factors that causally influence choice formation and the subjective attribution of the resulting choice.

On this basis, the present framework distinguishes three dimensions. The first is where objective intervention occurs within the choice–action–outcome chain. The second is the level—Decision, Action, or Outcome—at which self-attribution is preserved after that intervention. The third is whether that self-attribution takes the form of a feeling of agency grounded in sensorimotor cues or a judgment of agency incorporating context, beliefs, and inference. Forcing is a structure that arranges these dimensions asymmetrically, thereby producing a dissociation between objective influence and experienced authorship (Fig. 1).

## III. Engineering Agency Cues Through Forcing

### III.I. What Does Forcing Change?

Not every aspect of a magic performance that uses forcing can be reduced to the manipulation of agency. Trust in the performer, attention, memory, probability judgments, cold reading, and information asymmetry also contribute to the success of a performance. Nor is forcing a single psychological mechanism; it is a family of techniques that direct a choice or outcome by different means (Pailhès et al., 2020). Nevertheless, when forcing is organised around the spectator's own choice, the extent to which the spectator continues to attribute the decision, action, and outcome to themselves is relevant to the experience of illusory choice and control (Pailhès & Kuhn, 2020; Pailhès et al., 2021). In this restricted sense, forcing can be analysed as the engineering of agency cues.

A selection force intervenes in choice formation. In the Position Force, the spatial position of a card makes a particular card more likely to be selected, with reachability proposed as one contributing factor (Pailhès & Kuhn, 2021). Here, participants compare multiple genuine alternatives, commit to one of them, and execute the corresponding action themselves. What changes is choice probability, not the post-choice action–outcome contingency itself. Selection forces therefore provide a particularly clear case in which objective influence over choice may coexist with preserved subjective choice authorship.

An outcome force intervenes in action–outcome contingency. In the Criss-Cross Force, a spectator cuts a deck of cards at a freely chosen position, but the position of the cut does not determine the card that is ultimately revealed. Nevertheless, many participants fail to notice this structure and experience control over the outcome (Pailhès & Kuhn, 2020). What is primarily severed here is the Action–Outcome relation. This force therefore demonstrates a dissociation between Outcome-level Agency and objective causality, but does not by itself provide direct evidence of Decision-level Agency.

An interpretive force intervenes in the mapping between a response and its procedural meaning. In equivoque, whether the spectator's response is treated as choosing or eliminating an item is determined retrospectively so that a predetermined outcome is reached. Participants nevertheless report a sense of choice, freedom, and agency (Pailhès et al., 2021). Similarly, in choice blindness, a mismatch may go undetected when an outcome different from the participant's actual selection is presented, and reasons may subsequently be generated for the presented choice (Johansson et al., 2005).

These three forms show that agency is not simply present or absent at a single point. Selection forces bias choice probability while leaving intact cues that can support the Self–Choice relation. Outcome forces sever Action–Outcome contingency while leaving the choice and action intact. Interpretive forces alter Response–Meaning mapping while preserving the response itself. Differences between forcing techniques must therefore be described not only in terms of success rates, but also in terms of which relation within the choice–action–outcome chain is manipulated.

### III.II. How Are Agency Cues Manipulated?

The conclusion to be drawn from forcing is not simply that external manipulation produces an illusory experience of freedom. The agency genuinely retained by spectators can coexist with causal relations that have been altered by the performer. In a selection force, spectators compare candidates and point to, name, or grasp one of them. These actions are self-generated. Moreover, the decision process itself—'I compared the candidates and committed to one'—provides a cue supporting Decision-level Agency. Thus, even when the environment biases choice probability, cues capable of supporting authorship binding remain available. In an outcome force, the action performed by the spectator is genuinely self-generated, but the causal contingency between that action and the outcome is severed by the performer. If the procedure nevertheless unfolds naturally and the outcome remains temporally and semantically consistent with the immediately preceding action, cues supporting Outcome-level Agency may remain. In an interpretive force, the spectator genuinely generates the response, whereas what that response 'meant' is retrospectively determined by the performer. Here, the contextual interpretation underlying the judgment is reconstructed while agency over the local action is preserved.

This structure can also be explained through the distinction between feeling of agency and judgment of agency. So long as spectators speak, reach, and manipulate objects, a sensorimotor feeling of agency is likely to arise for those local actions. In addition, the prospective cue that they compared the options and committed to one themselves supports decision authorship. Against this background, the judgment that 'this choice originated from me' is constructed by integrating the performer's explanation, the structure in which the options were presented, consistency with the outcome, and awareness of manipulation.

The distinctive feature of forcing, therefore, is not merely the persistence of experienced agency despite external manipulation. Rather, agency genuinely retained by the spectator can serve as a scaffold for attributing even the manipulated relation to one's own choice or action. Forced yet free is not a simple opposition between illusion and reality; it is a combination of partially genuine agency and externally constructed causal relations.

### III.III. Why Does Forcing Look Like Mind Reading?

For a mind-reading effect to arise, it is not sufficient for the means by which the performer could know the choice to be hidden from the spectator. The spectator must also experience themselves as the author of the choice. If the spectator instead experiences the choice as having been imposed by the performer—'the performer made me choose this card'—the prediction can be explained as mere guidance rather than extraordinary mind reading. Conversely, if the choice is entirely random and not attributed to the spectator themselves, the structure of 'having my mind read' is weakened. Only when the spectator experiences the choice as 'my choice'—one that the performer should not have been able to know—can the performer's apparent knowledge of that choice constitute mind reading.

Forcing therefore does not merely engineer the success of a prediction. On the present account, it preserves cues supporting authorship binding, through which a choice can be constituted as 'mine', while concealing the fact that the choice could be predicted or controlled by the performer. It is this dual structure that may underpin the psychological effect of forcing.

## IV. Extending Forcing into XR

### IV.I. Embodied Space Shapes Choice Formation

The theoretical value of XR does not lie in its capacity to reproduce magic in three-dimensional space. Rather, XR makes it possible to continuously manipulate body-centred spatial relations while preserving both the semantic content of the options and their substantive availability for choice.

Distance, direction relative to the body midline, offset from the centre of the visual field, relation to the dominant hand, reachability, object orientation, required movement amplitude, motor cost, spatial audio, and appearance timing are not merely perceptual salience cues. They organise what is readily perceived, which actions are prepared, and which objects appear as 'available options'. In XR, multiple options can remain both formally and substantively available while the relative strength with which each object affords a particular action can be manipulated.

Consistent with this possibility, manipulating the visual salience of healthy products in a VR supermarket has been reported to alter choice without producing a clear difference in the decision-making experience (Blom et al., 2021). Because that study did not directly measure SoA or authorship binding, it does not constitute evidence for forced yet free. It does, however, suggest that changes in behavioural choice produced by environmental manipulation need not correspond one-to-one with changes in subjective decision experience.

More broadly, this proposal is consistent with existing research on self-experience in XR. In immersive video, a media format within VR, viewpoint-directed agency—the ability to select viewing direction through head rotation—can be preserved even when environment-directed agency, or the capacity to intervene in the recorded environment, is constrained (Toida, 2026b). Nevertheless, self-location can still be established at the recorded viewpoint. Likewise, presentness—the experience that another person is 'here, now' −may arise even in the absence of real-time co-presence (Toida, 2026c). It has further been reported that changing filming distance alone can alter Spatial Presence, Self-Location, Social Presence, Possible Actions, and related measures without changing system-level interaction capability (Toida et al., 2026).

These findings are not evidence for forcing. They are, however, methodologically aligned with the present argument insofar as they use the boundary conditions of XR to dissociate experiential components that ordinarily covary. The present paper extends this dissociation from presence and self-location to choice authorship. The viewpoint-directed/environment-directed agency distinction used in the earlier work and the present concept of Decision-level Agency, however, concern different axes. The former describes the target with respect to which action possibilities or control are available; the latter asks which decisions are experienced as having been initiated and settled by the self. The latter cannot be directly inferred from the former.

### IV.II. Biasing Choice While Preserving Authorship

In stage forcing, the performer's gesture, speech, gaze, card position, tempo, and other cues jointly direct choice. It is therefore difficult to isolate which cues affect selection probability and which preserve choice authorship. In XR, the relation between the body and the options can itself become a design variable. 'Ease of choosing' can be shaped, without issuing an instruction, by moving an object slightly closer, placing it on the dominant-hand side, shifting it towards the centre of the visual field, orienting it so that it is easier to reach, or changing the movement amplitude required to reach it.

Consider, for example, two virtual objects that are both genuinely selectable. If only one is positioned closer to the body, on the side of the dominant hand, and within reach using minimal arm movement, that object may be more likely to be chosen. Yet the user still perceives both objects, compares them, reaches out voluntarily, and selects one. The XR system is not making the choice on the user's behalf. Both options remain available, while commitment and motor execution remain intact. Cues capable of supporting Decision-level and Action-level Agency would therefore remain available. At the same time, choice probabilities are biased by the body-centred structure of the environment.

This would constitute an extension into egocentric space of the forced-yet-free structure produced by the stage-based Position Force. By introducing a gradient in compatibility with bodily parameters such as distance, reachability, posture, and motor cost, the action affordance associated with one object can be made stronger than that associated with another.

In this sense, XR is not merely an experimental apparatus for testing forcing. It is a medium in which the agency cues that magicians' forcing has bundled together in practice can be decomposed and recomposed as body-centred spatial relations. Crucially, 'an option exists', 'the option is physically selectable', 'one option is relatively easier to select', and 'the choice is attributed to oneself' are not equivalent conditions. By manipulating them separately, XR can reveal the conditions under which authorship binding is preserved or breaks down.

### IV.III. Forcing, AI, and XR

Forcing is not confined to the study of magic. It has been shown that forcing techniques derived from performance magic can be incorporated into goal-directed AI interaction, with participants maintaining relatively high levels of reported agency even when an AI steers the conversation towards a predefined target (Wen et al., 2026). This finding provides important empirical evidence that objective influence and experienced agency can coexist beyond stage magic. The forcing implemented in that study, however, was primarily conversational. The AI directed the decision process through question-driven narrowing, reframing, ambiguity, fallback strategies, and related techniques, whereas VR served as the task environment designed by the participant.

By contrast, the XR forcing proposed here operates through a different mechanism. The source of influence is not the conversation itself but the spatial relation between the body and the options. In other words, the environment alters the distribution of choices not through conversational content but through the spatial location of each option, its reachability, and the amount of movement required to select it.

The novelty of the present account therefore does not lie in being the first to identify a relation between forcing and agency. Research on magicians' forcing has demonstrated illusory control and freedom (Pailhès & Kuhn, 2020; Pailhès et al., 2021), and research on goal-directed AI has likewise shown the coexistence of forcing and reported agency (Wen et al., 2026). The contribution of the present paper is to decompose different forms of forcing as interventions along the choice–action–outcome chain, map them onto Decision-, Action-, and Outcome-level Agency, feeling of agency, judgment of agency, and Self–Choice authorship binding, and then extend this structure to egocentric spatial design.

### IV.IV. Theoretical and Ethical Implications

This framework shows that interactivity and autonomy cannot be treated as equivalent in XR. The fact that a user can move their own hand, manipulate an object, and produce an intended outcome supports Action-level and Outcome-level Agency. It does not, however, establish that the choice of which object to select was formed independently of environmental design. Conversely, the mere fact that the environment changed choice probability does not establish that Decision-level Agency was lost. The user may genuinely have compared the alternatives, committed to one, and initiated the action. Evaluating XR design therefore requires choice influence and choice authorship to be measured separately.

This distinction is also relevant to human–computer interaction (HCI) research, where the extent to which users experience themselves as producing outcomes through an interface has been examined in terms of agency. Intentional binding has also been introduced into HCI as a means of implicitly assessing interface-mediated agency (Coyle et al., 2012). Frameworks have further connected input modality, system feedback, computer assistance, human–computer joint action, and related factors to SoA research (Limerick et al., 2014). These lines of work primarily concern the relation between action execution and system outcome. Intentional binding can capture temporal/causal integration within that local action–outcome relation. It cannot, however, by itself reveal whether the choice underlying the selected option is experienced as self-generated. When XR intervenes in choice formation itself, authorship binding must therefore be treated as an independent object of analysis.

The distinction also has important ethical implications. In commercial, educational, or medical XR, if a user's choices are biased without the user recognising the guidance and are nevertheless experienced as self-determined, merely presenting multiple options in a formal sense is insufficient to support autonomy. High subjective authorship is not evidence of ethical legitimacy. In some cases, it may instead indicate that the source of influence is difficult to recognise.

At the same time, choice architecture cannot be eliminated altogether. The position, distance, ordering, and salience of options are inevitably subject to design. Choice architecture may also be directed towards user benefit, as in evacuation guidance, rehabilitation, or learning support. Evaluation should therefore consider not only whether influence exists, but also its purpose, benefits, substantive option availability, reversibility, transparency, contestability, and the possibility of withdrawal.

### IV.V. Scope Conditions and Falsifiability

The scope of the present argument is subject to several limitations. First, the paper does not reduce magic performance as a whole to agency, but focuses on forcing in which the spectator's own choice plays a central role in the performance. Second, authorship binding is not proposed as the same kind of psychophysical phenomenon as intentional binding. In the present account, authorship binding is a conceptual construct for describing the self-attributive relation between a choice and the self; no claim is made that a dedicated behavioural or implicit index has been established. Third, the argument does not claim that choice influence and decision authorship are partially independent across all forcing techniques or all choices. Authorship binding may break down when external influence is sufficiently strong or when the source of influence is explicitly recognised. Fourth, the present account does not claim that XR necessarily produces stronger forcing effects than those produced in real-world environments. The theoretical value of XR lies in the relative ease with which body-centred spatial relations can be manipulated independently, allowing option availability, motor ease, behavioural selection, awareness, and authorship—which ordinarily covary—to be experimentally dissociated.

The central claims of the present account are falsifiable. If objective choice influence uniquely determines decision authorship such that, once the magnitude of the bias in choice probability is controlled, authorship judgments decline to the same extent regardless of forcing technique, residual agency cues, action fluency, commitment, or outcome structure, then the agency-cue redistribution account proposed here would not be supported.

Likewise, if spatial manipulations of distance, reachability, motor cost, and related variables in XR always produce an equivalent reduction in authorship whenever they alter the choice distribution, such that behavioural influence and decision authorship cannot be experimentally dissociated, the proposed spatial extension of forced yet free would likewise not be supported.

## V. CONCLUSION: Toward Engineered Agency

This paper has extended the question of agency upstream, from intentional binding to choice authorship. Intentional binding is an important phenomenon for characterising the process through which action and outcome are linked into a single temporal/causal sequence. Yet it cannot by itself explain why a particular choice is experienced as one's own decision. Accordingly, authorship binding is conceptualised as the relation between a decision and the self as its subjective author. This does not denote a new temporal-perceptual phenomenon or implicit measure. Rather, it describes a constructive relation in which the resulting choice can be attributed to the self even when choice formation has been subject to external causal intervention.

Magicians' forcing provides a boundary case that exposes this relation. Selection forces preserve comparison and commitment while biasing choice probability. Outcome forces preserve self-generated action while severing action–outcome contingency. Interpretive forces preserve the response itself while altering its meaning. Forcing does not eliminate agency entirely; it selectively preserves, substitutes, and redistributes agency cues. Objective influence over choice and subjective authorship can therefore coexist without contradiction. Forced yet free does not mean a choice that is free from external influence. It describes a state in which, even though choice formation is environmentally directed, the decision is constructed as originating from the self.

XR can extend this structure into body-centred space. It can keep multiple choice options substantively available while strengthening the action affordance associated with one option relative to the others by manipulating distance, direction, reachability, posture, and motor cost. A user can still genuinely compare the alternatives, reach out, and make the choice, while the probability of that choice is biased by environmental structure. XR is therefore not merely an apparatus for reproducing forcing, but a conceptual medium in which choice influence can be dissociated from choice authorship, Action-level Agency from Decision-level Agency, and causal binding from authorship binding. If intentional binding opened the question, 'How are my action and its consequence bound together?', the next question exposed by forcing is: 'How is my decision bound to me as its author?'

## ACKNOWLEDGMENT

The author is grateful to those whose questions, suggestions, and conversations shaped the choices behind this paper. If you have reached this sentence, you have already made one more. I leave the question of its authorship to you.